\documentclass[
aps,%
12pt,%
final,%
notitlepage,%
oneside,%
onecolumn,%
nobibnotes,%
nofootinbib,%
noshowpacs,%
centertags]%
{revtex4}
\begin{document}
\selectlanguage{english}
\title{Detection of Giant Radio Pulses from the Pulsar PSR
B0656+14\footnote{Translated by G.~Rudnitskii} }
\author{\firstname{A.~D.}~\surname{Kuzmin}}
 \email[]{akuzmin@prao.psn.ru}
 \affiliation{ Pushchino Radio Astronomy Observatory, Astrospace Center, Lebedev Physical Institute, Russian
  Academy of Sciences, Pushchino, Moscow region, 142290 Russia }
%
\author{\firstname{A.~A.}~\surname{Ershov}}
 \email[]{ershov@prao.psn.ru}
 \affiliation{ Pushchino Radio Astronomy Observatory, Astrospace Center, Lebedev Physical Institute, Russian
  Academy of Sciences, Pushchino, Moscow region, 142290 Russia }
%
\begin{abstract}

A\textbf{\emph{bstract:}} Giant pulses (GPs) have been detected
from the pulsar PSR B0656+14. A pulse that is more intense than
the average pulse by a factor of 120 is encountered approximately
once in 3000 observed periods of the pulsar. The peak flux density
of the strongest pulse, 120 Jy, is a factor of 630 higher than
that of the average pulse. The GP energy exceeds the energy of the
average pulse by up to a factor of 110, which is comparable to
that for other known pulsars with GPs, including the Crab pulsar
and the millisecond pulsar PSR B1937+21. The giant pulses are a
factor of 6 narrower than the average pulse and are clustered at
the head of the average pulse. PSR B0656+14 along with PSR
B0031--07, PSR B1112+50, and PSR J1752+2359 belong to a
distinctive group of pulsars in which GPs have been detected
without extremely strong magnetic field on the light cylinder.


\textbf{\emph{Received:}} February 3, 2006

\textbf{\emph{Original Russian Text:}} Pis'ma v Astron. Zh., 2006,
v.32, pp.650-654

\textbf{\emph{English Text:}} Astronomy Letters, 2006, v.32,
pp.???

\end{abstract}
\maketitle
\newpage
\section{Introduction}

Giant pulses (GPs) are short-duration flare-like increases in the
intensities of individual radio pulses from pulsars. They received
the historical name 'giant' for the Crab pulsar PSR B0531+21,
which was discovered by Staelin and Sutton (1970) as a result of
the detection of very strong single pulses. Such pulses were
subsequently called giant ones.

This rare phenomenon was observed previously only in ten of the
more than 1500 known pulsars: PSR B0031--07 (Kuzmin et al. 2004;
Kuzmin and Ershov 2004), PSR J0218+4232 (Joshi et al. 2004), PSR
B0531+21 (Staelin and Sutton 1970), PSR B0540--69 (Johnston and
Romani 2003), PSR B1112+50 (Ershov and Kuzmin 2003), PSR
J1752+2359 (Ershov and Kuzmin 2005), PSR B1821--24 (Romani and
Johnston 2001), PSR J1823--3021(Knight et al. 2005), PSR B1937+21
(Wolszczan et al. 1984), and PSR B1957+20 (Joshi et al. 2004).
Seven of them belong to the group of pulsars that also radiate in
the high-energy part of the electromagnetic spectrum.

We have searched for and detected giant pulses from the eleventh
pulsar, PSR B0656+14. This pulsar is also known as an intense
source of optical, soft ultraviolet, X-ray, and gamma-ray pulse
emission (Koptsevich et al. 2001; Shibanov et al. 2005).

\section{Observations and data reduction}

The observations were performed with the BSA radio telescope at
the Pushchino Radio Astronomy Observatory (Astrospace Center,
Lebedev Physical Institute, Russian Academy of Sciences) with an
effective area at zenith of $15\,000~\mathrm{m^2}$. One linear
polarization was received. We used a 128-channel receiver with a
channel bandwidth of 20 kHz. The frequency of the first (highest
frequency) channel was 111.870 MHz. The time reference during the
observations was implemented using a GPS receiver.

The observations were carried out in the mode of recording
individual pulses. Since BSA is a transit telescope, the duration
of one observing session is limited to three minutes. In one
session, 520 periods of the pulsar were observed.

At the first stage of our search for and detection of GPs, from
April 30 through May 30, 2005, the sampling interval was 2.56 ms
at the time constant $\tau = 3$~ms. We carried out 26 observing
sessions containing 11\,648 periods of the pulsar. After the
detection of the first GPs, the observations were performed from
May 31 through October 2, 2005, with a sampling interval of
0.8192~ms at the time constant $\tau = 1$~ms to increase the time
resolution. In this mode, we carried out 86 observing sessions
containing 44\,720 periods of the pulsar.

Subsequently, all records were examined, de-noised, and
compensated for the dispersion delay in the interstellar medium.
In the derived time series, we found pulses that exceeded a preset
threshold level of the signal-to-noise ratio and determined their
amplitudes, durations, energies, and phases. We checked whether
pulses belonged to the pulsar by a check determination of the
dispersion measure from the maximum of the signal-to-noise ratio.

Figure 1 shows a sample record of one observing session (August 8,
2005). One GP with a peak flux density that is a factor of 630
higher than the average peak flux density of regular pulses hidden
in noise is distinguished in the record of a sequence of 520
pulsar periods.

The detected GPs are not the result of scintillations. The
scintillation time scale in the interstellar medium
(de-correlation time) for a pulsar with the dispersion measure $DM
= 13.977~\mathrm{pc~cm^{-3}}$ at the frequency of our
observations, 111 MHz, is $\sim 1$ min for diffractive
scintillations and $\sim 1\,000$ days for refractive
scintillations (Shishov et al. 1995). This is considerably longer
than the duration of the observed GPs (no more than several
milliseconds).

To obtain the session-averaged pulse profile and match these
profiles in phase for all observing sessions, we used the
ephemeris of this pulsar from the ATNF database (Manchester et al.
2005), including the dispersion measure $DM =
13.977~\mathrm{pc~cm^{-3}}$.

\section{Results}

In all 112 observing sessions (56\,368 pulsar periods), we
detected 52 pulses with the detection threshold at a
signal-to-noise ratio S/N > 5 (approximately one pulse in 1000
observed periods), including 19 pulses with S/N > 5.5.

We determined the GP flux density by comparison with the average
pulse of this pulsar with the measured flux density. The flux
density of the average (period-averaged) pulse $S_{AP} = 15$~mJy
for all 112 observing sessions was measured using the reference
source 3C 452; its flux density at 111 MHz was taken to be $S =
91$~Jy. The average pulse width is $w^{0.5}_{AP} = 30$~ms. The
peak flux density of the average pulse is $S^{peak}_{AP} =
S_{AP}P/w^{0.5}_{AP} = 190$~mJy.

The peak flux density of the strongest pulse, $S^{peak}_{GP} =
120$~Jy, exceeds that of the average pulse by a factor of 630. A
pulse that is a factor of 120 more intense than the average pulse
is encountered approximately once in 3000 observed periods of the
pulsar. The measured GP duration is $w^{0.5}_{GP} = 5$~ms. The
dispersion broadening is $\Delta t_{DM} = 1.7$~ms and the time
constant of the receiver, 1~ms, does not distort significantly the
pulsar GP shape.

Figure 2a shows the strongest observed GP together with the
average pulse for all 112 observing sessions.

The GP brightness temperature is
\[T_{\rm B}=S\lambda ^2/2k\Omega,\]
where $\lambda$ is the wavelength of the received radio emission,
$k$ is the Boltzmann constant, and $\Omega$ is the solid angle of
the radio-emitting region. Taking the size of the radio-emitting
region to be $l \le cw^{0.5}_{GP}$ and the distance to the pulsar
to be $d = 0.76$ kpc (Manchester et al. 2005), we obtain $T_B \ge
10^{26}$~K for $w^{0.5}_{GP} = 5$~ms and $S^{peak}_{GP} = 120$~Jy.

The energies of the strongest GP and the average pulse are,
respectively, $E_{GP} = S^{peak}_{GP} w^{0.5}_{GP} = 600$~Jy~ms
and $E_{AP} = S_{AP} P = 5.5$~Jy ms, where P is the pulsar period.
Thus, the energy of the strongest GP exceeds that of the average
pulsar pulse by a factor of 110.

Figure 3 shows the cumulative distribution of the ratio of the GP
energy to the average pulse energy $F(E_{GP}/E_{AP})$, where $F$
is the fraction of the pulses with energies above $E_{GP}/E_{AP}$.
This distribution was constructed from 34 GPs with energies that
are higher than the average pulse energy by a factor of 30 or more
and closely corresponds to a power law $F \propto (E_{\rm
GP}/E_{\rm AP})^\alpha$, which is typical of the GP intensity
distribution for known pulsars. The power-law index is $\alpha =
-2.0~\pm~0.3$~.

For $(EGP/EAP) < 30$ the ratio S/N < 5 and the distribution is
masked by noise. For this region, the dotted line indicates a
possible log-normal intensity distribution $\lg~F =
\lg~\exp(-(a(E_{\rm GP}/E_{\rm AP})^2~)$~,which is typical of
ordinary pulses (Cognard et al.1996). The parameter $a$ was chosen
to be 0.08; at this value, the Gaussian distribution passes
through our pulsar pulse detection threshold S/N = 5 at
$E_{GP}/E_{AP} \simeq 30$.

The phases of the observed GPs are shown in Fig. 2b. The GP
position is stable within the average pulse and GPs are clustered
at the head of the average pulse. The phase difference between GPs
(with S/N > 5.5) and the centroid of the average pulse is
$\Phi_{\rm GP} - \Phi_{\rm AP} = -13.1~\pm~2.4$~мс.

\section{Discussion}

The detected GPs from the pulsar PSR B0656+14 as well as the
previously detected GPs from the pulsars PSR B0031--07, PSR
B1112+50, and PSR J1752+2359 have all the characteristic features
of classical GPs: the GP peak flux density and energy exceed
significantly those of the average pulse, the GP energy
distribution corresponds to a power law, the GP duration is
considerably shorter than that of the average pulsar pulse, and
the GP phases are localized in a narrow emission window of the
average pulsar pulse.

To clarify the question of whether GPs are peculiar to a certain
class of pulsars, which is important in searching for GPs and
analyzing their radio emission mechanism, we compared some
parameters of the 11 known pulsars with GPs. The table lists some
comparative parameters of these pulsars and known energy
characteristics of the GPs themselves.

The table columns contain the following data: $PSR$ is the pulsar
name, $\nu$ is the frequency at which GPs were observed, $P$ is
the pulsar period, $B_{LC}$ is the magnetic field strength on the
light cylinder, + or - in the next column mean the presence or
absence of high-energy X ray emission, and $E_{GP}/E_{AP}$ is the
increase in the GP energy relative to the average pulse energy.

GPs were first detected from the pulsars PSR B0531+21 and PSR
B1937+21 with extremely strong magnetic fields on the light
cylinder, $B_{LC} < 10^6$~G. As a result, it was assumed that GPs
were inherent in pulsars with extremely strong magnetic fields on
the light cylinder and the searches for GPs were aimed at such
objects. However, in the pulsars PSR B0031--07, PSR B0656+14, PSR
B1112+50, and PSR J1752+23, the magnetic fields on the light
cylinder are 4~--~5 orders of magnitude weaker and are the same as
those in most of the known ordinary pulsars.

Johnston and Romani (2003) believe that GPs are peculiar to
pulsars that also radiate in the high-energy X-ray part of the
spectrum. However, 5 of the 11 pulsars, i.e., almost half of the
known pulsars with GPs, do not meet this criterion.

Knight et al. (2005) believe that GPs are inherent in pulsars with
large rotational energy losses, $\dot E \propto P^{-3}\cdot \dot
P$. However, GPs have been detected in pulsars with $\dot E$ that
differ by six orders of magnitude and cannot be an indicator for
their search either.

Giant pulses are observed over a wide frequency range, from 40 MHz
to 8 GHz, in pulsars of various types with a wide range of
periods, from 1.5 ms to 1.6 s, with magnetic fields on the light
cylinder from 4 to $10^6$~G. Nevertheless, in the known pulsars
with GPs, the flare-like increase in the GP energy is comparable
in order of magnitude, $E_{\rm GP}/E_{\rm AP} \sim 20~-~200$~.

\section{Conclusions}

We have detected giant pulses from the pulsar PSR B0656+14. The
peak flux density of the strongest pulse is a factor of 630 higher
than that of the average pulse. The GP energy exceeds that of the
average pulse by up to a factor of 110, which is comparable to
that for classical pulsars with GPs, PSR B0531+21 and PSR
B1937+21. The emission energy distribution corresponds to a power
law with the index $\alpha = -2.0~\pm~0.3$. The giant pulses are
considerably narrower than the average pulse and are clustered at
its head.

We performed a comparative analysis of the GP parameters for known
pulsars. We showed that GPs are inherent to pulsars of various
types and that the flare-like increase in the radio emission
energy in known pulsars with GPs differs by no more than an order
of magnitude.

\begin{acknowledgments}

We wish to thank V.V.~Ivanova, K.A.~Lapaev, A.S.~Aleksandrov, and
M.M.~Chereshnev for help with the observations. This work was
supported by the Russian Foundation for Basic Research (project
no. 05-02-16415), the State contract 02.452.12.7052, and the
'Non-stationary Phenomena in Astronomy' Program of the Presidium
of the Russian Academy of Sciences.

\end{acknowledgments}
\newpage
\begin{figure}[t!]
\setcaptionmargin{5mm}
\onelinecaptionsfalse 
\includegraphics{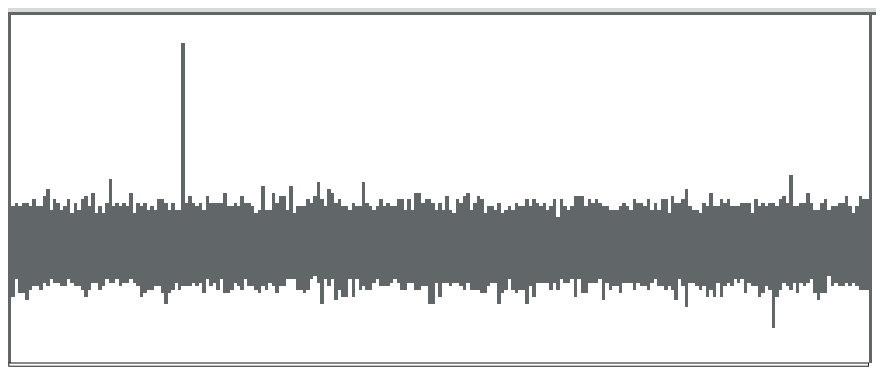}
\captionstyle{normal} \caption{ Sample record of the observing
session for August 8, 2005. Time (520 periods of the pulsar with a
time resolution of 0.819~ms) is along the horizontal axis;
intensity in arbitrary units is along the vertical axis. One giant
pulse is clearly seen.}
\end{figure}
\newpage
\begin{figure}[t!]
\setcaptionmargin{5mm}
\onelinecaptionsfalse 
\includegraphics{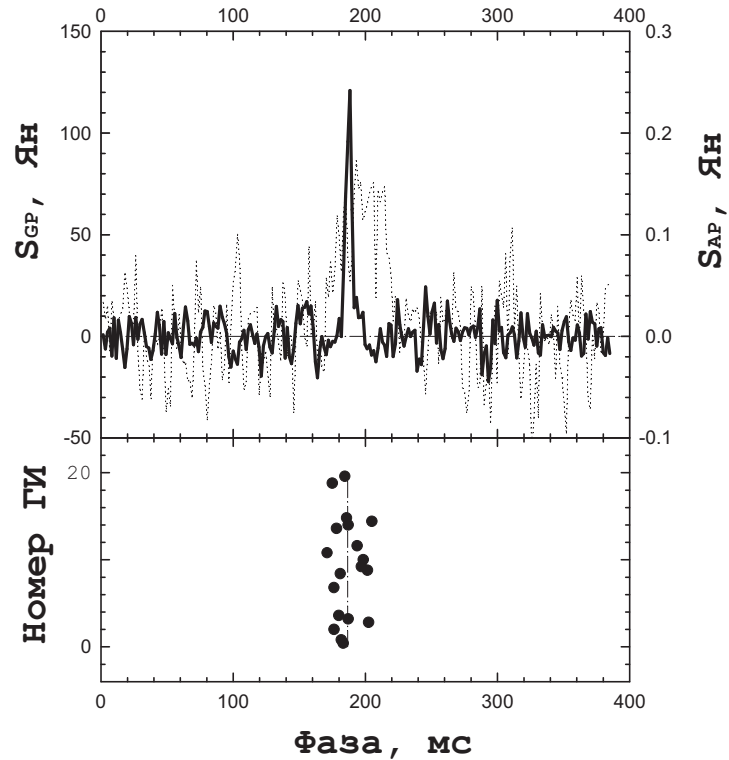}
\captionstyle{normal} \caption{ (a) The strongest observed giant
pulse (thick solid line) in comparison with the average pulsar
profile for all days of observations (thin dotted line). The peak
GP flux density is a factor of 630 higher then that of the average
pulsar pulse. For convenience, the average profile is magnified by
a factor of 500 and the flux densities of the giant and average
pulses are indicated separately on the left and right scales of
the vertical axis. (b) Phases of the giant pulses.}
\end{figure}
\newpage
\begin{figure}[t!]
\setcaptionmargin{5mm}
\onelinecaptionsfalse 
\includegraphics{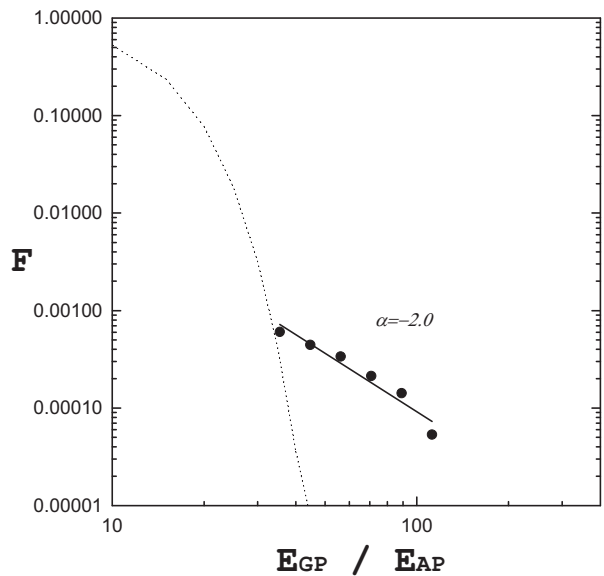}
\captionstyle{normal} \caption{ Cumulative distribution of the
giant-to-average-pulse energy ratio for the pulsar PSR B0656+14.
Here, F is the fraction of the pulses with energies above
$E_{GP}/E_{AP}$. The solid line indicates a power-law fit with the
index $\alpha = -2.0~\pm~0.3$~. The dotted line indicates a
possible version of the log-normal distribution.}
\end{figure}
\newpage
%
\begin{table}
\caption{
 }
\bigskip
\begin{tabular}{|l|c|c|c|c|r|c|}
\hline PSR & $\nu$~, MHz & $P$~, ms & $B_{\rm LC}$~, G & X-rays &
$\dfrac{E_{\rm GP}}{E_{\rm AP}}$ & References\footnote{1 -- Kuzmin
and Ershov, 2004; 2 -- Kuzmin et al., 2004; 3 -- Joshi et al.,
2004; 4 -- Kostyuk et al., 2003; 5 -- Johnston and Romani, 2003; 6
-- This paper; 7 -- Ershov and Kuzmin, 2003; 8 -- Ershov and
Kuzmin, 2005; 9 -- Romani and Johnston, 2001; 10 -- Knight et al.,
2005;
11 -- Soglasnov et al., 2004.}\\
\hline
B0031--07  &    40   &  942 &             7.0 &-- &  14 & 1 \\ 

           &   111   &      &                 &   &   8 & 2 \\ 
\hline
J0218+4232 &   400   & 2.32 & $3.2\cdot 10^5$ & + &  51 & 3 \\ 
\hline
B0532+21   &111-2228 & 33.4 & $9.8\cdot 10^5$ & + &  80 & 4 \\ 
\hline
B0540--69  & 1380    & 50.3 & $3.6\cdot 10^5$ & + &  -- & 5 \\ 
\hline
B0656+14   &     111 &  385 & $7.7\cdot 10^2$ & + & 110 & 6 \\ 
\hline
B1112+50   &     111 & 1656 &             4.2 &-- &  16 & 7 \\ 
\hline
J1752+2359 &     111 &  409 &              71 &-- & 200 & 8 \\ 
\hline
B1821--24  &    1517 & 3.05 & $7.4\cdot 10^5$ & + &  81 & 9 \\ 
\hline
J1823--3021&     685 & 5.44 & $2.5\cdot 10^5$ &-- &  64 &10 \\ 
           &    1400 &      &                 &   &  28 &  \\
\hline
B1937+21   &    1650 & 1.57 & $1.0\cdot 10^6 $& + &  60 &11 \\ 
\hline
B1957+20   &     400 & 1.60 & $3.8\cdot 10^5$ &-- & 129 & 3 \\ 
\hline
%
%
\end{tabular}
\end{table}

\newpage
\begin{thebibliography}{99}
%
\bibitem{cognard}
\refitem{article} I.~Cognard, J.A.~Shrauner, J.H.~Taylor,
S.E.~Thorsett, Astrophys. J. 457, L81 (1996).
%
\bibitem{ershov03}
\refitem{article;rusjou} A.A.~Ershov and A.D.~Kuzmin, Pis'ma
Astron. Zh. 29, 111 (2003). [Astron. Lett. 29,91 (2003)].
%
\bibitem{ershov05}
\refitem{article} A.A.~Ershov, A.D.~Kuzmin, Astron. Astrophys.
443, 593 (2005).
%
\bibitem{johnston}
\refitem{article} S.~Johnston, R.W.~Romani, Astrophys. J. 590, L95
(2003).
%
\bibitem{joshi}
\refitem{book} B.C.~Joshi et~al., \emph{Young Neutron Stars and
Their Environments, IAU Symp. 218}, (Eds. F.~Camilo,
B.M.~Gaensler, San Francisco: ASP, 2004), p. 319.
%
\bibitem{knight}
\refitem{article} H.S.~Knight, M.~Bailes, R.N.~Manchester,
S.M.~Ord, Astrophys. J. 625, 951 (2005).
%
\bibitem{koptsevich}
\refitem{article} A.B.~Koptsevich, G.G.~Pavlov, S.V.~Zharikov,
et~al., Astron. Astrophys. 370, 1004, (2001).
%
\bibitem{kostyuk}
\refitem{article;rusjou} S.V.~Kostyuk, V.I.~Kondrat'ev,
A.D.~Kuzmin et al., Pis'ma Astron. Zh. 29, 440 (2003)
[Astron.Lett. 29, 387 (2003)].
%
\bibitem{kuzmin04a}
\refitem{article} A.D.~Kuzmin, A.A.~Ershov, Astron. Astrophys.
427, 575 (2004).
%
\bibitem{kuzmin04b}
\refitem{article;rusjou} A.D.~Kuzmin, A.A.~Ershov, and
B.Ya.~Losovsky, Pis'ma Astron. Zh. 30, 285 (2004) [Astron. Lett.
30, 247 (2004)].
%
\bibitem{manchester}
\refitem{article} R.N.~Manchester, G.B.~Hobbs, A.~Teoh, M.~Hobbs,
Astron. J. 129, 1993 (2005).
%
\bibitem{romani}
\refitem{article} R.W.~Romani, S.~Johnston, Astrophys. J. 557, L93
(2001).
%
\bibitem{shibanov}
\refitem{article} Yu.A.~Shibanov, J.~Sollerman, P.~Lundqvist
et~al., Astron. Astrophys. 440, 693, (2005).
%
\bibitem{shishov}
\refitem{article;rusjou} V.I.~Shishov, V.M.~Malofeev,
A.V.~Pynzar', and T.V.~Smirnova, Astron. Zh. 72, 485 (1995)
[Astron. Rep. 39, 428 (1995)].
%
\bibitem{soglasnov}
\refitem{article} V.A.~Soglasnov, M.V.~Popov, N.~Bartel, et~al.,
Astrophys. J. 616, 439 (2004).
%
\bibitem{staelin}
\refitem{article} D.H.~Staelin, J.M.~Sutton, Nature 226, 69
(1970).
%
\bibitem{wolszchan}
\refitem{book} A.~Wolszczan, J.M.~Cordes, D.R.~Stinebring,
\emph{Millisecond Pulsars} (Eds. S.P.~Reynolds, D.R.~Stinebring,
Green Bank: NRAO, 1984), p. 63.
%
\end{thebibliography}
\end{document}